\documentclass[12pt]{article}
\usepackage{amssymb,graphicx} 
\usepackage{amsmath}
\usepackage{epsfig}

\def\e{{\rm e}}

\def\d{\partial}
\def\l{\left(}
\def\r{\right)}

\newcommand{\be}{\begin{equation}}
\newcommand{\ee}{\end{equation}}

\newcommand{\Tr}{{\rm Tr}}
\newcommand{\bg}{\begin{gather}}
\newcommand{\eg}{\end{gather}}

\begin{document}
\begin{center}
  {\Large\bf Gauge theory solitons on noncommutative
  cylinder} \\
\medskip  
S.V.~Demidov$^a$, S.L.~Dubovsky$^b$, V.A.~Rubakov$^b$, S.M.~Sibiryakov$^b$\\
\medskip
  {\small
$^a$
Moscow Institute of Physics and Technology,\\
Institutskii per., 9, Dolgoprudny, 141700 Moscow Region, Russia\\
   $^b$Institute for Nuclear Research of
         the Russian Academy of Sciences,\\  60th October Anniversary
  Prospect, 7a, 117312 Moscow, Russia
  }
\end{center}

\begin{abstract}
We generalize to noncommutative cylinder the solution
generation technique, originally suggested for gauge theories on
noncommutative plane. For this purpose we construct partial isometry
operators and complete set of orthogonal projectors in the algebra
${\cal A}_C$ of the cylinder, and an isomorphism between the free
module ${\cal A}_C$ and its direct sum ${\cal A}_C\oplus {\cal F}_C$
with the Fock module on the cylinder. We construct  explicitly the
gauge theory soliton and evaluate the spectrum of perturbations about
this soliton.
\end{abstract}

\section{Introduction.}
Recently, field theories on noncommutative (NC) spaces have attracted
considerable interest (see,
e.g. Refs. \cite{konechnyshwartz,harvey,nekrasovdouglas,arefeva} 
for recent reviews and references to
ealier work). One of the motivations is the fact that
NC Yang-Mills theory emerges as effective low-energy
description of string theory in background
$B$-field in a certain limit \cite{seibergwitten}. Of particular 
interest is the study of solitons in NC theories. These solitons
share many common properties with D-branes;
most notably, correct values of D-brane tensions are 
reproduced, and the spectrum of fluctuations about
NC solitons is 
in qualitative and quantitative agreement with the
spectrum of open strings in D-brane backgrounds \cite{Harvey:2000jt,Aganagic:2000mh}.

The best studied example of NC field theory is a theory on 
NC plane
(more generally, on noncommutative $R^{2n}$). The algebra
${\cal A}_P$ of
functions on NC plane is generated by two coordinates
$\hat{x}_1\equiv \hat{x}$
and $\hat{x}_2\equiv\hat{p}$ obeying the following commutation relation,
\[
[\hat{x},\hat{p}]=i\theta\;.
\]
This algebra is isomorphic to the algebra of operators in 
quantum mechanics of one degree of freedom.

In the limit of strong noncommutativity, $\theta\to\infty$ , solitons
of pure scalar theory are constructed in terms of projectors $P$ in
the algebra ${\cal A}_P$ \cite{gopakumar} which satisfy
\[
P*P=P\;,
\]
where $*$ is multiplication in ${\cal A}_P$.
For gauge theory on NC plane, 
solution generation technique was suggested
\cite{Aganagic:2000mh,Harvey:2000jb,harvey} 
which enables one to obtain new exact solutions starting from 
vacuum. This technique makes use of partial isometry
operators, which are elements $S$ of the algebra ${\cal A}_P$ satisfying the
following properties,
\be
\label{partial}
SS^+=1,\;S^+S=1-P\;,
\ee
where  $P$ is a projector.
Thus, solitons in NC gauge theory obtained by the
solution generation technique are again labeled by projectors in
the algebra ${\cal A}_P$.

As explained in Ref.  \cite{konechnyshwartz}, the 
solution generation technique is closely related to an
isomorphism between the free module over the algebra ${\cal
A}_P$  and the direct sum ${\cal F}_P\oplus {\cal A}_P$ of the
Fock module ${\cal F}_P$ 
(which is merely the Hilbert space of harmonic
oscillator states) and the free module. In the case of NC plane, 
however, the use of the
formal algebraic language of projective modules may appear
an unnecessary sophistication, since the
 solution generation technique, as presented originally 
\cite{Aganagic:2000mh,Harvey:2000jb,harvey}, is very transparent by itself.

The next simplest examples of NC spaces are NC
cylinder and torus.  Corresponding algebras will be denoted in what
follows by ${\cal A}_C$ and ${\cal A}_T$, respectively. In the case of
gauge theory on NC torus, it is problematic to construct a soliton in
terms of connection on ${\cal A}_T$ (to the best of our knowledge, no
explicit solitonic solution is known so far). 
One may still construct a NC gauge theory soliton as 
connection on ${\cal F}_T\oplus{\cal A}_T$,
 the direct sum of the Fock and free modules
\cite{Konechny:2001gi}. 
 This approach
enables one also to calculate the
spectrum of small perturbations about the soliton
\cite{Konechny:2001gi}, 
which is in agreement with the spectrum of strings in 
 D0-D2 background, now in toroidally compactified 
space. However, ${\cal F}_T\oplus{\cal A}_T$ and ${\cal A}_T$ are not
isomorphic \cite{konechnyshwartz}, thus a connection on ${\cal
F}_T\oplus{\cal A}_T$ does not induce a connection on the free module
${\cal A}_T$, and no solution generation technique emerges.

In this paper we find and study exact classical solutions of 
$U(1)$ gauge theory on NC cylinder (see Ref. \cite{headrick} for the
description of scalar solitons on NC cylinder 
in the large-$\theta$ limit and Refs. \cite{chaichian, bak, hamanaka} 
for alternative 
approaches to NC cylinder). In particular, 
we discuss whether
an analog of the solution generation technique exists in this
case. We find an affirmative answer to this question and present
explicit solitonic solution directly in terms of connection on the
algebra ${\cal A}_C$. Unlike the case of NC plane, 
the algebraic formalism described in
Ref. \cite{konechnyshwartz} proves  most adequate 
in gauge theory on NC cylinder. 

This paper is organized as follows.
In Section \ref{algebra} we describe the algebra ${\cal A}_C$ of functions on
the noncommutative cylinder as a subalgebra of the algebra
${\cal A}_P$ of NC plane. We introduce a basis in the Fock space
${\cal F}_P$ which is convenient for constructing
projectors and partial isometry operators in 
the cylinder algebra ${\cal A}_C$. We also describe connections
on the Fock and free modules of this algebra.
In Section \ref{solitons} we present the solution generation technique
for NC cylinder and give an explicit construction of gauge theory 
soliton as a connection 
on the free module ${\cal A}_C$. In Section \ref{spectrum} we study
the fluctuation spectrum about this soliton by making use of its representation
as a connection on the direct sum ${\cal A}_C\oplus{\cal F}_C$.

\section{Noncommutative cylinder.}
\label{algebra}
\subsection{Algebra, projectors, partial isometry operators}
In {\it commutative} geometry, the algebra of functions on a cylinder
of radius $R$ with coordinates $(x,p)$
may be defined as a subalgebra of functions on a plane, subject to
invariance under finite translations
\be
\label{trans}
A(x,p)= A(x+2\pi R,p)\;.
\ee
On NC plane with coordinates $(\hat{x},\hat{p})$,  generators of 
infinitesimal translations are inner 
derivations, 
\be
\label{derivative}
\frac{\partial\hat{A}}{\partial\hat{x}}=\frac{i}{\theta}[\hat{p},\hat{A}]\ ;\;\;
\frac{\partial\hat{A}}{\partial\hat{p}}=-\frac{i}{\theta}[\hat{x},\hat{A}]\;.
\ee
Consequently, Eq.~(\ref{trans}) is generalized to the
noncommutative case by defining the algebra ${\cal A}_C$ of functions on 
NC cylinder as a subalgebra of
the algebra ${\cal A}_P$, which  consists of functions
 commuting with the operator of 
finite translation along $x$-direction
\begin{equation}
\label{NCtrans}
\hat{A}\in{\cal A}_C:\;
\e^{-2\pi i R\hat{p}/\theta}\hat{A}\e^{2\pi i R\hat{p}/\theta}=\hat{A}\;.
\end{equation}
The generators of the algebra ${\cal A}_C$
are $\hat{p}$ and
$\exp(i\hat{x}/{R})$; clearly, the coordinate $\hat{x}$
does not belong to ${\cal A}_C$.
The condition  
(\ref{NCtrans}) is preserved by the operation of differentiation
(\ref{derivative}), so Eq. (\ref{derivative}) is still
a definition of the derivatives on NC cylinder. Note that the
differentiation with respect to $\hat{p}$ is not an inner derivation on
the algebra ${\cal A}_C$.

The algebra ${\cal A}_P$ may be viewed as an algebra of
operators acting in the Hilbert space of functions of one variable. 
In what follows it will be
convenient to work in $p$-representation. Then
every operator
$\hat{A}$ is uniqely determined by its kernel $A(p,p')$,
\[
\l\hat{A}\psi\r (p)=\int d 
p'
A(p,p')\psi(p')\;,
\]
where $\psi(p)$ is an arbitrary function.
The constraint (\ref{NCtrans})  implies that an operator belongs to the
subalgebra ${\cal A}_C$ iff its kernel satisfies the
following condition,
\[
A(p,p')=A(p,p')\exp\left(-2\pi i R(p-p')/\theta\right),  
\]
or, equivalently,
\be
\label{apq}
A(p,p')=\sum_{q\in {\mathbf Z}}a(p,q)\delta\left({Rp}/{\theta}-
{Rp'}/{\theta}-q\right)
\ee
with arbitrary coefficient functions $a(p,q)$ of one continuous and
one discrete variable.

To construct field theory on NC cylinder one needs to define a trace
$\Tr_C$ on the algebra ${\cal A}_C$. This trace is a noncommutative
generalization of the integral over cylinder. The trace cannot be 
simply the trace
$\Tr$ on the algebra ${\cal A}_P$. Indeed, the representation (\ref{apq})
makes it clear that trace $\Tr$ diverges for the elements of ${\cal
A}_C$.
This is not a surprise: in the commutative case one would
obtain  divergent results, if functions on cylinder were understood
as periodic functions on a plane, and integral over the cylinder were
defined as 
an integral over the whole plane.

Instead, one defines the integral on a cylinder as the integral of
periodic function over its period. The way to generalize this
definition to the noncommutative case is to recall that the algebra
${\cal A}_C$ is generated by two elements $\hat p$ and
$\e^{i\hat{x}/R}$, i.e., every element of this algebra has the form
\be
\label{2*}
\hat{A}=\sum_{q=-\infty}^{\infty}a(\hat{p},q)e^{iq\hat{x}/R}\;.
\ee
(in fact, the Fourier components
$a(\hat{p},q)$ coincide with the coefficient functions appearing in
the expression (\ref{apq}) for the kernel of the operator $\hat A$.)
One defines $\Tr_C \hat{A}$ as an integral of the zeroth
Fourier component of the operator $\hat A$,
\be
\Tr_C\hat{A}\equiv\int_{-\infty}^{\infty}a(p,0)dp\;.
\ee
It is straightforward to check that $\Tr_C$ has all defining
properties of the trace operation.

By making use of this trace one writes for the Weyl symbol of the
operator (\ref{2*}), 
\begin{equation}
\label{21}
f_{\hat A}(x,p)=\frac{1}{2\pi}\sum_{n}{}\int d\tau e^{-in{x}/{R}-i\tau {Rp}/
{\theta}}Tr_C\left[\hat{A}(\hat{x},\hat{p})e^{i(n{\hat{x}}/{R}+\tau
{\hat{p}R}/{\theta})}\right].
\end{equation}
Clearly, this symbol is periodic in $x$ with the period $2\pi R$.
The inverse transformation is
\[
\hat{A}(\hat{x},\hat{p})=\frac{1}{(2\pi)^2}\sum_m\int d\sigma 
e^{-i(m{\hat{x}}/{R}+\sigma{R\hat{p}}/{\theta})}\tilde{f}_{\hat A}(m,\sigma)
\]
where
\be
\label{fourier}
  \tilde{f}_{\hat A}(m,\sigma)=\int_0^{2\pi R}dx\int d
p
  e^{i(m{x}/{R}+\sigma{Rp}/{\theta})}f_{\hat{A}}(x,p) 
\ee
Now, it is straightforward to check the following relation between the
trace of an operator and the integral of its Weyl symbol  over the cylinder,
\[
  2\pi\theta Tr\hat{A}=\int_Cdxdpf_{\hat A}(x,p).
\]
This relation has the same form as in the case of NC plane, and is
precisely what is needed for constructing field theory on NC cylinder.

Let us now construct projectors and partial isometry operators in the
algebra ${\cal A}_C$ of NC cylinder. 
In the Hilbert space of
functions $\psi(p)$, it is convenient to choose an orthonormal
basis with specific properties. Namely, the elements of
this basis,  $|n,m\rangle$, are labeled by two numbers  
$n=0,1,2,\dots$ and $m=0,\pm 1,\pm 2,\dots$ The defining property
of this basis is the following transformation rule under the 
finite translation,
\be
\label{transmn}
\e^{2\pi i R\hat{p}/\theta}|n,m\rangle=|n,m+1\rangle\;.
\ee
Making use of Eq.~(\ref{NCtrans}) one finds that 
the latter property and orthonormality condition 
 imply that any element  of the
algebra ${\cal A}_C$ can be presented in the following form
\be
\label{Cmnl}
\hat{A}=\sum_{n,n',m,l}C_{n,n'}^{l}|n,m\rangle\langle m+l,n'|\;,
\ee
where $C^l_{nn'}$ are arbitrary coefficients. One can then
construct the following set of
projectors,
\be
\label{4*}
P_i=\sum_m|i,m\rangle\langle m,i|\;.
\ee
They have the general form of Eq.~(\ref{Cmnl}), {\it i.e.}, belong to
the algebra of NC cylinder.
These projectors are orthogonal to each other,
\[
\label{PiPj}
P_iP_j=\delta_{ij}P_i
\]
and form a complete set in the space of projectors in the 
algebra ${\cal A}_C$.
To see the latter property, let us
check that there is no non-zero projector
$P\in{\cal A}_C$ such that for all $i$
\be
\label{PPi}
PP_i=0
\ee
For any projector $P$ and any basis function
$|i,m\rangle$
one has
\be
\label{Pmnmn}
 \left|\left|P|i,m\rangle\right|\right|^2=\Tr\l P |i,m\rangle\langle
m,i|\r \geq 0\;.
\ee
If $P\in {\cal A}_C$ then, according to Eqs. (\ref{NCtrans}) and
(\ref{transmn}), the trace in Eq.~(\ref{Pmnmn}) does not
depend on $m$. Consequently, one may write 
\[
\left|\left|P|i,m\rangle\right|\right|^2=\frac{1}{2N+1}\sum_{m=-N}^{m=N} \Tr\l P
|i,m\rangle\langle i,m|\r\;.
\]
Because of Eq. (\ref{PPi}), the sum here tends to zero as
$N\to\infty$.
Hence, any projector 
$P\in {\cal A}_C$ satisfying eq.~(\ref{PPi})
 annihilates all basis functions $|i,m\rangle$ and, 
consequently, is equal to zero.

It is also straightforward to construct a set of partial isometry
operators, analogous to the shift operators on the plane,
\begin{equation}
\label{Cisometry}
  S^C_i=\sum_{m,n}^{}|n,m\rangle \langle m,n+i|\;,
\end{equation}
with the property
\[
  S^C_iS_i^{C+}=1,\]
\[ S_i^{C+}S^C_i=1-P_i\;.
\]      
Again, these operators have the general form (\ref{Cmnl}), so they
belong to the algebra ${\cal A}_C$.

The existence of the orthonormal basis
$|n,m\rangle$ with the property (\ref{transmn}) is far from being
obvious. Let us prove by construction that such basis indeed exists.
The translation property  (\ref{transmn}) implies that the basis
elements $|n,m\rangle$ have the following form in the 
$p$-representation
\be 
\label{nmmoment}
|n,m\rangle=\xi_{n+1}(p)\exp\left({2\pi i Rmp}/{\theta}\right)\;,
\ee
while orthonormality and completeness of the set $\{|n,m\rangle\}$ imply
\be
\label{ort}
\int dp\xi_n(p)\xi_{n'}(p)e^{2\pi i{Rp}
(m-m')/\theta}=\delta_{nn'}\delta_{mm'}
\ee
and
\begin{eqnarray}
\sum_{n,m}^{}\xi_n(p)\xi_n(p')e^{2\pi i {R(p-p')}m/{\theta}}
=\delta\left(p-p'\right),
\label{comp}
\end{eqnarray}
Now, let $\chi_n(\lambda)$ be an arbitrary
orthonormal basis on the interval $\lambda\in [-\pi,\pi]$ such that
\[
\chi_n(-\pi)=\chi_n(\pi)=0\;.
\]
Then, it is straightforward to check that functions 
\be
\label{genxi}
\xi_n(p)=\sqrt{R\over \theta}\int_{-\pi}^{\pi}d\lambda
\chi_n(\lambda)
\e^{2\pi i Rp\lambda/\theta}
\ee
have the desired properties (\ref{ort}) and (\ref{comp}). Thus any
auxiliary basis $\chi_n(\lambda)$ defines a basis $|n,m\rangle$
obeying the translation property (\ref{transmn}).

A relatively simple example of functions $\xi_n(p)$ 
is generated by the following choice of the auxiliary basis,
\[
\chi_n(\lambda)={1\over \sqrt \pi}\sin\left[{(\pi+\lambda) }(n+1)/2\right]
\]
The corresponding functions $\xi_{n}(p)$ are
\begin{equation}
\label{ourxis}
\xi_{2k}=\sqrt{2R\over\theta}\frac{k}{\pi}\frac{\sin\pi {Rp}/{\theta}}
{\left({Rp}/{\theta}\right)^2-k^2},
\end{equation}
and
\be
\label{ourxis1}
\xi_{2k+1}=\sqrt{2R\over\theta}\frac{(k+\frac{1}{2})}{\pi}\frac{\cos\pi{Rp}/{\theta}}
{\left({Rp}/{\theta}\right)^2-(k+\frac{1}{2})^2},
\ee
where $k=0,1,2,\dots$ Making use of Eqs. (\ref{nmmoment}), (\ref{ourxis})
and (\ref{ourxis1}), one can construct projectors (\ref{4*}) and
partial isometry operators (\ref{Cisometry}) explicitly.

\subsection{Modules and endomorphisms of NC cylinder.}
Noncommutative analogues of vector bundles and connections
(=gauge fields) on these bundles are projective modules
and derivatives on these modules (see, e.g.,
Refs. \cite{konechnyshwartz,nekrasovdouglas} for reviews). 
In the previous subsection we
constructed a complete orthonormal set of projectors in the algebra
${\cal A}_C$. This construction implies, in analogy to the case of NC
plane, that an arbitrary projective module over ${\cal A}_C$ is a direct sum
of a certain number of free and Fock modules. Let us describe
these two modules in some detail.

Elements of the (right) free module are elements of the 
algebra ${\cal A}_C$ itself, and the algebra acts on this module by right
multiplication. An arbitrary connection on this module
can be represented in the form 
\be
\label{freegauge}
\nabla_i=\d_i+iu_i\;,
\ee
where 
\be
\label{fderiv}
\d_1=\frac{\partial}{\partial\hat{x}}\;,\;\;\d_2=
\frac{\partial}{\partial\hat{p}}
\ee
and $u_i$ belongs to the
algebra $End_{{\cal A}_C}({\cal A}_C)$ of endomorphisms of the free
module, i.e., similarly to the case of NC plane,
$u_i$
is an arbitrary element of the 
algebra ${\cal A}_C$, acting on the free module by left
multiplication.
The curvature $F_{12}$ is defined in the usual way,
\[
F_{12}=[\nabla_1,\nabla_2]\;.
\]
Clearly, the ``canonic'' connection $\d_i$ on the free module has
zero curvature.

Elements of the (right) Fock module ${\cal F}_C$ of NC cylinder
are ``bra''-vectors  $\langle f|$ of the Fock module of NC plane with
the standard action of operators $\hat{A}\in {\cal A}_C\subset{\cal A}_P$,
\[
\langle f|\hat{A}=(\hat{A}^+|f\rangle)^+
\]
This module has a constant curvature connection
$\nabla_{i}^{0}$, with the following components
\begin{eqnarray}
\nabla_1^0(\langle f|)=-\frac{i}{\theta}\langle f|\hat{p}\nonumber\\
\nabla_2^0(\langle f|)=\frac{i}{\theta}\langle f|\hat{x}\nonumber\;.
\end{eqnarray}
The curvature of this connection is equal to
\[
F_{12}=[\nabla_1^0,\nabla_2^0]=-\frac{i}{\theta}\;.
\]
An arbitrary connection on the Fock module
can be represented in the form 
\[
\nabla_i= \nabla_i^0+iz_i\;.
\] 
where $z_i$ belongs to the 
algebra $End_{{\cal A}_C}({\cal F}_C)$ of endomorphisms of the module
${\cal F}_C$,
i.e., gauge fields  $z_i$ on the Fock module have the following form
\be
\label{fockgauge}
  z_i=\sum_{n}^{}z_i(n)\e^{2\pi i nR\hat{p}/{\theta}},
\ee
where $z_{i}(n)$ are some numbers. Note that unlike the case of 
NC plane, where one could have only constant gauge fields on the Fock
module, $z_i$ are not constant in general.

In what follows we will encounter also the direct sum  $ {\cal F}_C \oplus
{\cal A}_C $ of the Fock and free modules. To set the notations
let us write its elements as a column
\[
\l  
\begin{array}{c}
   \langle f| \\
    \hat{A}
  \end{array}
\r
\;.
\] 
An arbitrary connection on such a module can be written in the form
\begin{equation}
\nabla_i=\tilde{\nabla}_i^0+i
\l
   \begin{array}{cc}
    z_i      & \psi_i \\
    \psi_i^* & v_i
   \end{array}
\r
\;,
\label{endom}
\end{equation}
where
\be
\label{sumcon} 
\tilde{\nabla}_1^0=
\l
\begin{array}{cc} 
-\frac{i}{\theta}\hat{p}&0\\
  0                     &\d_1
\end{array}
\r\;\;\;\;
 \tilde{\nabla}_2^0=
\l
\begin{array}{cc}
\frac{i}{\theta}\hat{x}&0\\
  0                    &\d_2
\end{array}
\r
\ee
Here $v_i\in End_C ({\cal A}_C)$, while 
$\psi$ and $\psi^{*}$ are homomorphisms
from the free module ${\cal A}_C$ to the Fock module ${\cal F}_C$ 
and backwards. These homomorphisms can be parametrized
by vectors $\langle\psi_i|$ and $|\psi_i\rangle$
in the following way,
\begin{eqnarray}
  \psi_i(\hat{A})&=&\langle\psi_i|\hat{A}\nonumber\\
  \psi_i^*(\langle f|)&=&\sum_m  \e^{2\pi i m
  R\hat{p}/\theta}|\psi_i\rangle\langle f|
\e^{-2\pi i m R\hat{p}/\theta}\;,
\label{vectors}
\end{eqnarray}
where  $\hat{A}\in {\cal A}_C$ and $\langle f|\in {\cal F}_C$. 
\section{Solution generation technique on NC cylinder.}
\label{solitons}
Let us now consider  $U(1)$ gauge theory on NC cylinder.
The action is
\[
S=\frac{\pi\theta}{2g^2}\int dt\;Tr_CF^{\mu\nu}F_{\mu\nu}\;,
\]
where $\mu,\nu=0,1,2$.
The field
strength $F_{\mu\nu}$ is defined in the usual way,
\[
F_{\mu\nu}=[\nabla_\mu,\nabla_\nu]\;,
\]
where $\nabla_i$
 ($i=1,2$) are defined in Eq.~(\ref{freegauge}) and
\[
\nabla_0=\d_t+iu_0\;.
\]
Gauge field components $u_\mu$ are elements of the algebra ${\cal A}_C$
(acting on the free module by left multiplication). In general, these
fields are time dependent. In this section we will search for 
{\it static} solutions
of the field equations, in the gauge
\[
u_0=0\;.
\]
Then, the static field equations have the usual form
\be
\label{stat1}
[\nabla_i,[\nabla_i,\nabla_j]]=0\;.
\ee
The purpose of this section is to obtain solution generation
technique in the $U(1)$ gauge theory on NC cylinder, which will be 
analogous to the technique suggested in
Ref.~\cite{Aganagic:2000mh,Harvey:2000jb,harvey} 
for NC plane.

Let us first  recall how the solution generation technique works on NC
plane \cite{Aganagic:2000mh,Harvey:2000jb,harvey}.
There, due to the fact that derivatives (\ref{fderiv}) are in
fact inner derivations on the algebra ${\cal A}_P$ 
(cf. Eq.~(\ref{derivative})), one can introduce the following operators
\[
C_1={i\over\theta}\hat{p}+iu_1\;\;\;C_2=-{i\over\theta}\hat{x}+iu_2\;.
\]
Then, static field equations (\ref{stat1}) take the following 
form
\be
\label{stat2}
[C_i,[C_i,C_j]]=0\;.
\ee
The distinction between Eqs. (\ref{stat1}) and (\ref{stat2}) is that
the former involves connections on the free module of the algebra
${\cal A}_P$, while the latter involves the elements of the algebra itself.
Now, the key observation is that if elements $C_i^0$ solve
Eq.~(\ref{stat2}) and $S$ is an arbitrary element of the
algebra ${\cal A}_P$ satisfying
\[
SS^+=1\;,
\]
then operators
\be
\label{sgt}
\tilde{C}_i=S^+C_i^0S
\ee
also solve Eq.~(\ref{stat2}). If $S$ is a unitary
operator, i.e.,
\[
S^+S=1\;,
\]
then the field configuration $\tilde{C}_i$ is just a gauge
transform of the configuration $C^0_i$. However, if $S$
is a partial isometry operator (see Eq.~(\ref{partial})) then
$\tilde{C}_i$ make a new solution of the field
equations. For instance, one may start from vacuum configuration,
with $u_i=0$, and obtain $n$-soliton solutions by making use of the
shift operators
\[
S_n=\sum_{m}^{}|m\rangle \langle m+n|\;.
\]
In particular, the one-soliton solution is 
\[
\tilde{C}_1=\frac{i}{\theta}S_1^+\hat{p}S_1~,~~~
\tilde{C}_2=-\frac{i}{\theta}S_1^+\hat{x}S_1\;,
\] 
or, in terms of gauge fields
\begin{eqnarray}
\label{10*}
u_1={1\over i}\l S_1^+\d_1S_1-{i\over\theta}P_0\hat{p}\r\\
u_2={1\over i}\l S_1^+\d_2S_1+{i\over\theta}P_0\hat{x}\r
\label{10**}
\end{eqnarray}
where $P_0=|0\rangle\langle 0|$.

It is easy to see that a naive attempt to generalize this
construction to the case of NC cylinder fails. Indeed, operators $C_i$
do not belong to the algebra of NC cylinder. Still one may try to make
use of the embedding of the algebra ${\cal A}_C$ into the algebra
${\cal A}_P$ and write blindly the same formula (\ref{sgt}), with
partial isometry operator $S^C$ belonging now to ${\cal A}_C$ (see
Eq.~(\ref{Cisometry})).  However, it is straightfroward to see that
the resulting gauge fields do not belong to the
subalgebra ${\cal A}_C$, so the whole construction fails.

To obtain the solution generation technique on NC cylinder, let us
generalize the formalism \cite{konechnyshwartz} relating the solution 
generation
technique on NC plane to the isomorphism between modules ${\cal A}_P$
and ${\cal A}_P\oplus {\cal F}_P$. Let us make use of the basis 
introduced in 
Section 2.1. 
Then we define the isomorphism $\Lambda:{\cal A}_C\to {\cal A}_C\oplus
{\cal F}_C$ 
as follows,
\be
\label{10+}
  \Lambda:\hat{A}\to
\l
\begin{array}{c}
\langle 0,0|\hat{A} \\
S^C_1\hat{A}
\end{array}
\r\;
,
\ee
where $S^C_1$ is defined in Eq.~(\ref{Cisometry}).
The inverse map is 
\[
  \Lambda^{-1}:
\l
\begin{array}{c}
\langle f|\\
\hat{A}
\end{array}
\r
\to S_1^{C+}\hat{A}+\phi\l\langle f|\r\;,
\]
where 
\[
\phi\l\langle f|\r=\sum_{m}{}|0,m\rangle \langle f|\e^{-2\pi 
iRm\hat{p}/\theta}\;.
\]
Let us mention the following useful property of the operation $\phi$,
\be
\label{useful}
\phi(\langle f|\hat{A})=\phi\l\langle f|\r\hat{A}
\ee
for any $\hat{A}\in {\cal A}_C$.

Now, let us note that the connection (\ref{sumcon}) on the direct sum
${\cal A}_C\oplus {\cal F}_C$ satisfies Eq.~(\ref{stat1}).
Then, by making use of the isomorphism $\Lambda$ one may define the
following connection $\nabla_i$ on 
the free module ${\cal A}_C$
\begin{equation}
\label{gennablas}
\nabla_i=\Lambda^{-1}\circ\tilde\nabla_i^0\circ
\Lambda.
\end{equation}
In the case of NC plane,
similar construction is equivalent to the solution
generation technique \cite{konechnyshwartz}. Here, too, the connection 
(\ref{gennablas}) obeys the field equation (\ref{stat1}), i.e., it
corresponds to a soliton. Exact multi-soliton solutions  on NC
cylinder may be obtained by repeating this procedure.

Let us now explicitly 
calculate the
gauge field for one soliton solution on NC cylinder, 
resulting from this procedure.
Namely, let us consider the action of the connection $\nabla_i$ on
arbitrary element $\hat{A}\in {\cal A}_C$. Making use of the property
(\ref{useful}) one finds
\be
\label{nablas}
\nabla_i(\hat{A})=\l \d_i+S_1^{C+}\d_iS_1^C-
\phi(\langle 0,0|X_i)\r\hat{A}
,
\ee
where
\[
X_1={i\over\theta}\hat{p},\;\;\; X_2=-{i\over\theta}\hat{x}\;.
\]
Since $X_1\in {\cal A}_C$, we may further simplify the expression for
$\nabla_1$ by using again the 
property (\ref{useful}). We arrive at the following
result for the $u_1$-component of the gauge field
\[
u_1= {1\over i}\l S_1^{C+}\d_1S_1^C-
{i\over\theta}P_0\hat{p}\r\;.
\]
This would coincide with Eq. (\ref{10*}) valid in the case of NC plane, if
the partial isometry operator in Eq. (\ref{10*}) were $S_1^C$.

Since $\hat{x}$ is not an element of ${\cal A}_C$, an expression like
(\ref{10**}) cannot be written  for $u_2$. Yet the formula
(\ref{nablas}) simplifies in a basis of a certain class.
Namely, one may rewrite the
$\phi$-term for  $u_2$ in the following way,
\begin{eqnarray}
\phi(\langle 0,0|X_2)=\phi(\langle 0,0|P_0X_2)
=\phi(\langle 0,0|[P_0,X_2])+\phi(\langle 0,0|X_2)P_0\nonumber\\
=P_0[P_0,X_2]+\sum_{m,l} |0,m\rangle\langle 0,0|X_2|0,l\rangle\langle 0,l+m|\;,
\label{simplif}\;
\end{eqnarray}
where in the second line we made use of the fact that
$[P_0,X_2]\in{\cal A}_C$.
Now, it is straightforward to check that if the basis is of
the form (\ref{genxi}) with functions $\chi_n(\lambda)$ of definite
parity (this is the case for the choice (\ref{ourxis}), (\ref{ourxis1}))
 then
\[
\langle n,0|X_2|n,l\rangle=0
\]
for all $n,l,$ and the last term in Eq. (\ref{simplif}) drops out.
In this case one finds the following expression for the
$u_2$-component
of the gauge field,
\[
u_2= {1\over i}\l S_1^{C+}\d_2S_1^C+P_0\d_2 P_0\r
\]
The curvature may be expressed in terms of the
projector $P_0$  without
reference to any particular basis and is equal to 
\begin{equation}
\label{curvature}
F_{12}=[\nabla_1,\nabla_2]= -\frac{i}{\theta}P_0\;,
\end{equation}
which is clear from Eqs. (\ref{gennablas}) and (\ref{sumcon}). 

The energy of the soliton is
\[
 E=-\frac{\pi\theta}{g^2}Tr_{C}F^{12}F_{12}=\frac{\pi}{g^2\theta}Tr_{C}P_0=
\frac{\pi}{g^2\theta}.\]
This coincides with the energy of the soliton on NC plane.

The
curvature $F_{12}$ and its Weyl symbol are not gauge-invariant
quantities. Yet the shape of the soliton (in a certain gauge) may be
of some interest.
The Weyl symbol of the operator $F_{12}$ obtained with the basis
(\ref{ourxis}), (\ref{ourxis1}),
is shown in Fig.1 (see
Appendix for the sketch of the calculations leading to this graph).
\begin{figure}[!h]
\begin{center}
\epsfig{file=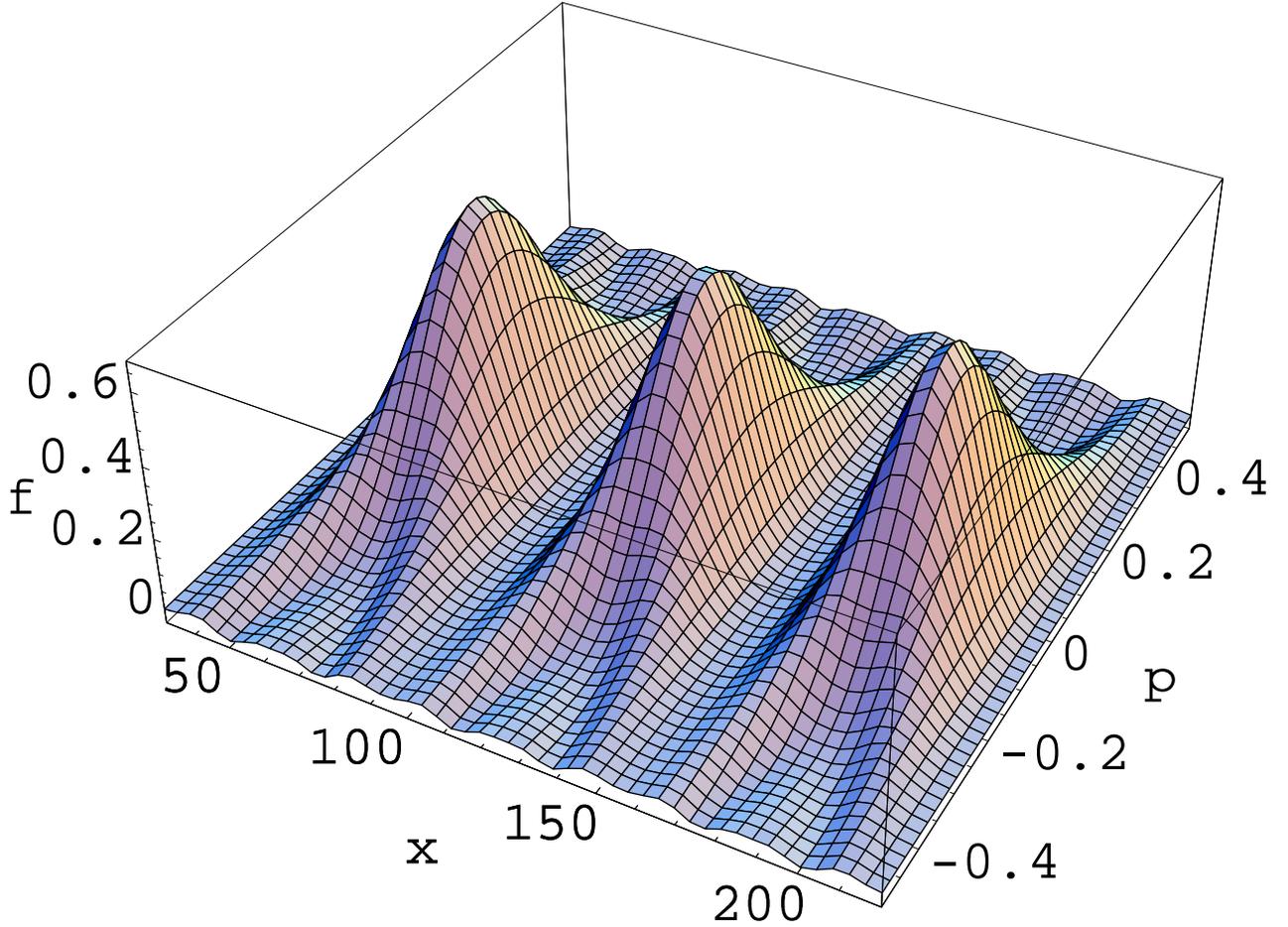,height=14.cm,width=17.cm}
\caption{The Weyl symbol of  $F_{12}$ on the plane $(x,p)$ for
$R=10$ and $\theta=3$. Periodicity in $x$-coordinate means that one
deals with the soliton on the cylinder.}
\end{center}
\end{figure}

\section{Fluctuation spectrum.}
\label{spectrum}
To study the spectrum of perturbations about the soliton found in
Section 3 it is convenient, following Ref. \cite{Konechny:2001gi}, 
to make use of the isomorphism $\Lambda$, Eq. (\ref{10+}), 
and rewrite the action for perturbations in the
soliton background in
terms of connections on the direct sum ${\cal A}_C\oplus {\cal F}_C$.
The first step is to define the trace on the endomorphisms of
this module. For endomorphism of general form given by
Eq.~(\ref{endom}) one defines
\[
 \Tr_{{\cal A}\oplus {\cal F}}
\l
  \begin{array}{cc}
  z_i      & \psi_i \\
  \psi_i^* & v_i
  \end{array}
\r
=\Tr_C(P_0 z_i)+\Tr_C v_i\;.
\] 
It is straightforward to check that this trace is an image of the
trace $\Tr_C$ under isomorphism $\Lambda$.

We make use of the following notation,
\[
\Tr \psi_2\psi_1^*=\langle \psi_1|\psi_2\rangle\;,
\]
where vectors $|\psi\rangle$ are introduced in eq.~(\ref{vectors}).
Then the action for perturbations 
in the soliton background in the gauge $A_0=0$ is 
\[
S=\frac{\pi\theta}{2g^2}\Tr_{\cal{A}\oplus {\cal F}} \int dt(-2[\d_t,\nabla_i]^2+
[\nabla_i,\nabla_j]^2),
\]
where   
$\nabla_i=\tilde{\nabla}_i^{0}+i\delta A_i$. 
Here 
$\delta A_i$ is the Hermitan matrix of fluctuations,
\[
\delta A_i=\l
\begin{array}{cc}
     z_i  &\psi_i\\
 \psi^*_i& v_i
\end{array}
\r
\]
The quadratic part of the action can be written as a sum of contributions
of four decoupled sectors. The first sector includes fields $v_i$ (in
the string language this sector describes $(2,2)$-strings ---
open strings with both ends on the $D2$-brane that fills up the
cylinder), the second sector includes fields $z_i$
(corresponding to the $(0,0)$-strings with both ends on
the $D0$-brane (soliton)) and the third sector includes off-diagonal
components $\psi$, $\psi^*$ ($(0,2)$-strings with one end on the
$D2$ and another end on the $D0$-brane).

The quadratic action in the first sector takes the following form
\[
 S_v=\frac{\pi\theta}{2g^2}\Tr_C\int dt\left(2(\partial_0v_i)^2-2
(\partial_1v_2-\partial_2v_1)^2\right)\;.
\]
This coincides with the quadratic part of the action for $U(1)$ gauge theory
on NC cylinder without soliton, in the $u_0=0$ gauge. At the quadratic level,
commutative and noncommutative theories coincide, so the spectrum in
the $(2,2)$-sector is just the spectrum of Abelian gauge field on the
cylinder.

In the $(0,0)$-sector one has, at the quadratic level,
\[
S_z={2\pi\theta \over g^2}\int dt\sum_{n=0}^{\infty} \left[|\d_t
z_i(n)|^2-4\pi^2{R^2\over\theta^2}n^2|z_1(n)|^2\right]\;,
\]
where $z_i(n)$ are defined in Eq.~(\ref{fockgauge}), and we made use
of the Hermiticity condition
\[
z_i(-n)=z_i(n)^*\;.
\]
This action describes a tower of massive modes localized on the
soliton, which are labeled by $z_1(n)$,
$n=1,2,\dots$, with masses
\be
\label{winding}
m_n=2\pi n{R\over\theta}\;,
\ee
and two  massless moduli $z_i(0)$ with $n=0$ 
corresponding to the position of
the soliton on NC cylinder. All modes $z_2(n)$ with $n\neq 0$ are
unphysical and are eliminated by the Gauss' law constraint, which has the
following form in the $(0,0)$-sector,
\[
\d_0\d_iz_i=0\;
\]
It is worth noting that the masses 
(\ref{winding}) are proportional to the radius of the cylinder, so
the physical massive modes  $z_1(n)$ 
correspond to the winding modes in the string theory language.

Finally, in the off-diagonal sector one arrives at the following action,
\be
\label{02action}
S_\psi={2\pi\theta\over g^2}\int dt\l
\Tr\d_0\psi_i^*\d_0\psi_i-{1\over\theta}\l
 ||a|\psi_-\rangle-a^+|\psi_+\rangle||^2-\langle\psi_-|\psi_-\rangle+
\langle\psi_+|\psi_+\rangle\r\r
\ee
where we introduced new fields
\begin{eqnarray}
\psi_1=\frac{i}{\sqrt{2}}(\psi_--\psi_+)\nonumber\\
\psi_2=\frac{1}{\sqrt{2}}(\psi_-+\psi_+),\nonumber
\end{eqnarray}
and creation/annihilation operators
\begin{eqnarray}
a^+=\frac{1}{\sqrt{2\theta}}(\hat{x}+i\hat{p})\nonumber\\
a=\frac{1}{\sqrt{2\theta}}(\hat{x}-i\hat{p})\nonumber\;.
\end{eqnarray}
Expanding
\[
|\psi_{\pm}\rangle=\sum_{n=0}^{\infty}t_n^{\pm}|n\rangle,
\]
where $|n\rangle$ are oscillator eigenstates, and substituting
this expression into the action, we obtain the following expression
for the potential term in Eq.~(\ref{02action}),
\[
-\frac{4}{\theta}(\sum_{n=1}^{\infty}|t_{n+1}^-\sqrt{n}-t_{n-1}^+\sqrt{n+1}|^2-
|t_0^-|^2)\;.
\]
This mass matrix is diagonalized by the following transformation,
\begin{eqnarray}
\varphi_n=\frac{1}{\sqrt{2n+1}}(t^-_{n+1}\sqrt{n}-t^+_{n-1}\sqrt{n+1})
\nonumber\\
\eta_n=\frac{1}{\sqrt{2n+1}}(t^-_{n+1}\sqrt{n+1}+t^+_{n-1}\sqrt{n})\nonumber\;,
\end{eqnarray}
yielding the result
\[
S_{\psi}=\frac{2\pi\theta}{g^2}\int
dt\left[|\partial_0t_0^-|^2+{1\over\theta}
|t_0^-|^2+|\d_0\eta_0|^2+
\sum_{n=1}^{\infty}\l|\partial_0\varphi_n|^2+|\partial_0\eta_n|^2-
\frac{2n+1}{\theta}|\varphi_{n}|^2\r\right].
\]
The Gauss' law constraint in this sector has the form
\[
\partial_0(\nabla^0_i\psi_i)=0,
\]
implying that all $\eta_n $ are unphysical.   
Among the physical fields
there is one tachyonic mode $t_0^-$ of mass squared
$-\frac{1}{\theta}$, and the tower 
of massive fields $\phi_n$, $n=1,2,\dots,...$.
with the mass spectrum
$m_{n}^{2}=\frac{2n+1}{\theta}$.

All spectra calculated in this Section agree with those in string
theory on a cylinder. This again confirms the interpretation  of NC
solitons as D-branes.
\section{Discussion}
To summarize, the key point of 
this paper is the explicit construction of the
orthonormal 
basis $|n,m\rangle$ in the Hilbert space of functions of one variable.
Elements of this basis transform according to the rule (\ref{transmn})
under the action of the translation operator. 
This
property enables one to construct a complete set of orthogonal
projectors (\ref{4*}) and corresponding set of partial isometry operators
(\ref{Cisometry}) in the algebra ${\cal A}_C$ of NC cylinder. 
By making use of this partial isometry
operators, we found an isomorphism (\ref{10+})  
between the free module over the algebra ${\cal
A}_C$  and direct sum ${\cal F}_C\oplus {\cal A}_C$ of the
Fock module ${\cal F}_C$ and the free module. This construction
leads to the generalization of the solution generation technique
to the case of NC cylinder. We explicitly described the
gauge field of the one-soliton solution and calculated the spectrum of
small perturbations about this solution.

In this concluding section let us discuss another implication of the
existence of the basis  $|n,m\rangle$. Namely, in Ref. \cite{grossnekrasov},
the following peculiar property of gauge theory on NC plane was
pointed out. Consider $U(N)$ gauge theory with a certain 
$N$ on NC plane. Then, for {\it any} natural $K$,
there exists a vacuum in this
theory, such that the theory above  
this vacuum is identical to $U(K)$ gauge theory above trivial vacuum. 
An interpretation
of this property given in
Ref. \cite{grossnekrasov} is that the number of colors $N$ emerges as a
superselection parameter, labeling separate sectors of the quantum
Hilbert space of an NC gauge theory.

Technically, this property is due to the fact that in the case of NC
plane, there exists an isomorphism between sums of free modules $\sum_1^N
{\cal A}_P$ and $\sum_1^K
{\cal A}_P$ with any $N$ and $K$. Gauge theories in these sums are
$U(N)$ and $U(K)$-theories, respectively, so this isomorphism induces
a connection corresponding to the $U(K)$ vacuum in the $U(N)$ gauge
theory.

In the case of NC torus, this property does not seem to hold as
there is no isomorphism between direct sums $\sum_1^N
{\cal A}_T$ and $\sum_1^K
{\cal A}_T$ with different $N$ and $K$ \cite{konechnyshwartz}.
Let us demonstrate that the properties of the basis  $|n,m\rangle$ imply
that NC cylinder is similar to NC plane in this respect, 
{\it i.e.} $U(N)$ gauge theories with different $N$ emerge on NC cylinder
as theories above different backgrounds in a single gauge theory.

For the sake of simplicity let us discuss how $U(2)$ gauge theory
emerges in $U(1)$ gauge theory (the generalization to
other $N$ and $K$ is straightforward). First, let us construct an
isomorphism between modules ${\cal A}_C$ and ${\cal A}_C\oplus {\cal
A}_C$.
Elements of the direct sum  ${\cal A}_C\oplus {\cal
A}_C$ are columns
\[
\l
\begin{array}{c}
\hat{A}_1\\
\hat{A}_2
\end{array}
\r\;,
\]
where $\hat{A}_1$ and $\hat{A}_2$ 
are elements of the algebra ${\cal A}_C$. Let us define a
map  $\Lambda:{\cal A}_C\to {\cal A}_C\oplus {\cal 
A}_C$ as follows,
\[
\Lambda:
\hat{A}\to \l
\begin{array}{c}
E \hat{A}\\
O \hat{A}
\end{array}
\r\;,
\]
where the operators $E$ and $O$ are
\begin{eqnarray}
E&=&\sum_{m,n}|n,m\rangle\langle m, 2n|\nonumber\\
O&=&\sum_{m,n}|n,m\rangle\langle m, 2n+1|\nonumber\;.
\end{eqnarray}
It is straightforward to check the following properties of these
operators
\begin{eqnarray}
OO^+=EE^+=O^+O+E^+E=1\nonumber\\
EO^+=OE^+=0\;.\nonumber
\end{eqnarray}
By making use of these properties, one checks
that the inverse map $\Lambda^{-1}:{\cal 
A}_C\oplus {\cal A}_C\to {\cal A}_C$ is
\[
\Lambda^{-1}:\l
\begin{array}{c}
\hat{A}_1\\
\hat{A}_2
\end{array}
\r
\to
E^+\hat{A}_1+O^+\hat{A}_2\;.
\]
Hence, $\Lambda$ is an isomorphism.
In complete similarity to
Section 3, one constructs the connection 
on the module ${\cal A}_C$, 
\[
\nabla_i=\Lambda^{-1} \circ \d_i\circ\Lambda
\] 
induced from the
vacuum connection $\d_i$ on the module ${\cal A}_C\oplus
{\cal A}_C$. It is
straightforward to calculate the $U(1)$ gauge field corresponding to
the connection $\nabla_i$,
\be
\label{u1}
A_i=E^+\d_iE+O^+\d_iO\;.
\ee
By construction, the curvature of this gauge field is
equal to zero, and $U(1)$ theory on NC cylinder
in the background (\ref{u1}) coincides with $U(2)$ theory on NC
cylinder in trivial background. 

Thus, any given $U(N)$ gauge theory on either NC plane or NC cylinder
has an infinite set of vacua labeled by $K=1,2,\dots$. Above each
vacuum, this theory is equivalent to $U(K)$ gauge theory above its
trivial vacuum. It is natural to wonder whether these vacua correspond
to different supeselection sectors, or are merely different phases of one
and the same theory, like, say, degenerate vacua with different vev's
in (commutative) scalar theories. The theory on NC cylinder is, in
principle, adequate to address this question: in the latter case,
there should exist a domain wall configuration (not necessarily
solution) of {\it finite energy}, separating  different vacua.
It would be interesting either to construct such a configuration, or
prove that there  is none.
\section*{Acknowledgements}
The authors are indebted to A.Konechny for useful discussions.
This work has been supported in part by Russian Foundation for Basic
Research, grants 02-02-17398, 02-02-06514, 02-02-06515, 
and Swiss Science Foundation grant
7SUPJ062239.  The work of S.Dubovsky has
been supported in part by INTAS grant YS 2001-2/128. The work of S.S. has
been supported in part by INTAS grant YS 2001-2/141.

\section*{Appendix. Weyl symbol of
$F_{12}$.}
Let us calculate the Weyl symbol of the
curvature (\ref{curvature}) obtained with the
basis given by  (\ref{ourxis}), (\ref{ourxis1}).
For the sake of simplicity let us  set $R=\theta=1$ in this
calculation. 
Expression (\ref{4*}) implies that the kernel of the projector $P_0$
has the following form
\[
P_0(p,p')=\sum_{m}^{}\xi_1(p)\xi_1(p')\exp{(2\pi im(p-p'))}.
\]
Hence, for coefficient functions $a(p,q)$ (cf. eq.~(\ref{apq})) one finds
\[
a(p,q)=\xi_{1}(p)\xi_{1}(p-q)
\]
Making use of Eq. (\ref{21}) one
writes 
\[
\tilde{f}_{P_0}(m,p)=2\pi a(p-m/2,-m),
\]
where the Fourier transform $\tilde{f}_{P_0}(m,p)$ of the Weyl symbol is
defined in eq. (\ref{fourier}).
Then the expression for the Weyl symbol itself is
\begin{eqnarray}
f(x,p)&=&\sum_{n}^{}\xi_{1}(p-n/2)\xi_{1}(p+n/2)\exp{(-inx)}\nonumber\\
&=&\frac{\cos{(2\pi p)}}{2\pi^{2}p}\sum_{n=-\infty}^{n=\infty}
\cos{(nx)}\left(\frac{1}{n^2-(2p+1)^2}-\frac{1}{n^2-(2p-1)^2}\right)
\nonumber\\
&+&\frac{1}{2\pi^{2}p}\sum_{n=-\infty}^{n=\infty}
\cos{(nx)}\left(\frac{(-1)^{n}}{n^2-(2p+1)^2}-
\frac{(-1)^{n}}{n^2-(2p-1)^2}\right)
\label{series}
\end{eqnarray}
Evaluation of the series in eq. (\ref{series}) results in the
following explicit expression for the Weyl symbol $f_{P_0}(x,p)$
(we restore the dependence on $R$ and $\theta$ in these final formulae)
\begin{enumerate}
\item{$x\in [0,\pi R]$}.
\begin{eqnarray}
f_{P_0}(x,p)&=&
\frac{1}{\theta}
\left[\frac{
\cos{\left(2\pi Rp/\theta\right)}}
{2\pi^{2}Rp/\theta}
\left(\frac{\pi}{2Rp/\theta-1}
\frac{\cos{\left[\left(\pi-x/R\right)
\left(2Rp/\theta-1\right)\right]}}
{\sin{\left(\pi\left(2Rp/\theta-1\right)\right)}}\right.\right.
\nonumber\\
&-&\left.\left.
\frac{\pi}{2Rp/\theta+1}\frac{\cos{\left[\left(\pi-x/R\right)
\left(2Rp/\theta+1\right)\right]}}
{\sin{\left(\pi\left(2Rp/\theta+1\right)\right)}}\right)\right.\nonumber\\
&+&\left.\frac{1}{2\pi^{2}Rp/\theta}
\left(\frac{\pi}{2Rp/\theta-1}\frac{\cos{\left(2xp/\theta-x/R\right)}}
{\sin{\left(\pi\left(2Rp/\theta-1\right)\right)}}\right.\right.
\nonumber\\
&-&
\left.\left.
\frac{\pi}{2Rp/\theta+1}\frac{\cos{\left(2px/\theta+x/R\right)}}
{\sin{\left(\pi\left(2Rp/\theta+1\right)\right)}}\right)
\right]\nonumber
\end{eqnarray}

\item{$x\in [\pi R,2\pi R]$}.
\begin{eqnarray}
f_{P_0}(x,p)&=&\frac{1}{\theta}\left[\frac{\cos{\left(2\pi Rp/\theta\right)}}
{2\pi^{2}Rp/\theta}
\left(\frac{\pi}{2Rp/\theta-1}\frac{\cos{\left[\left(\pi-x/R\right)
\left(2Rp/\theta-1\right)\right]}}
{\sin{\left(\pi\left(2Rp/\theta-1\right)\right)}}\right.\right.\nonumber\\
&-&\left.\left.
\frac{\pi}{2Rp/\theta+1}\frac{\cos{\left[\left(\pi-x/R\right)
\left(2Rp/\theta+1\right)\right]}}
{\sin{\left(\pi\left(2Rp/\theta+1\right)\right)}}\right)\right.\nonumber\\
&+&\left.\frac{1}{2\pi^{2}Rp/\theta}
\left(\frac{\pi}{2Rp/\theta-1}\frac{\cos{\left[(2\pi-x/R)\left(2Rp/\theta-1\right)\right]}}
{\sin{\left(\pi\left(2Rp/\theta-1\right)\right)}}\right.\right.\nonumber\\
&-&\left.\left.
\frac{\pi}{2Rp/\theta+1}\frac{\cos{\left[(2\pi-x/R)\left(2Rp/\theta+1\right)\right]}}
{\sin{\left(\pi\left(2Rp/\theta+1\right)\right)}}\right)
   \right]\nonumber
\end{eqnarray}
\end{enumerate}
This function is not analytic at $x=\pi R$; however one can explicitly
check that $f_{P_0}$ and its derivative are continuous
at $x=\pi R$.

\end{document}